# RareCollab: an LLM-powered framework for multimodal reasoning in Mendelian disease diagnosis

Guantong Qi[1,2,3,*], Jiasheng Wang[2,3,*], Mei Ling Chong[4], Zahid Shaik[2,3], Shenglan Li[4], Shinya Yamamoto[2,3], Maura R.Z. Ruzhnikov[6,7], Devon E. Bonner[6,7], Jennefer N. Carter[6,7], Kevin S. Smith[7], Matthew T. Wheeler[5,8], Stephen B. Montgomery[7,9,10], Jonathan A. Bernstein[5], Sasidhar Pasupuleti[2,3], Undiagnosed Diseases Network, Pengfei Liu[4,†], Hu Chen[2,3,†], Zhandong Liu[2,3,†]

1. Graduate School of Biomedical Sciences, Program in Genetics & Genomics, Baylor College of Medicine, Houston, TX, 77030, USA
2. Department of Pediatrics, Jan and Dan Duncan Neurological Research Institute, Baylor College of Medicine, Houston, TX, 77030, USA
3. Jan and Dan Duncan Neurologic Research Institute, Texas Children's Hospital, Houston, TX, 77030, USA
4. Department of Molecular and Human Genetics, Baylor College of Medicine, Houston, TX, 77030, USA
5. Stanford Center for Undiagnosed Diseases, Stanford University, Stanford, CA, 94305, USA.
6. Department of Pediatrics, Division of Medical Genetics, Stanford University, Stanford, CA, 94305, USA.
7. Department of Pathology, Stanford University School of Medicine, Stanford, CA, 94305, USA.
8. Department of Medicine, Division of Cardiovascular Medicine, Stanford University, Stanford, 94305, CA, USA.
9. Department of Genetics, Stanford University School of Medicine, Stanford, CA, 94305, USA.
10. Department of Biomedical Data Science, Stanford University School of Medicine, Stanford, CA, 94305, USA.

* These authors contributed equally
† Correspondence: pengfeil@bcm.edu, hu.chen@bcm.edu, zhandonl@bcm.edu

## Abstract

Rare disease diagnosis increasingly relies on integrating genomic, phenotypic and transcriptomic evidence, yet these signals remain difficult to reconcile within a common interpretive framework. Here we present RareCollab, an LLM-powered framework for multimodal reasoning in Mendelian disease diagnosis that integrates more than 100 diagnostic evidence signals across DNA, RNA, phenotype, curated variant-level knowledge, and in-silico pathogenicity evidence. This design enables large language models to operate as calibrated, interpretable reasoning modules rather than as a single end-to-end ranker. We applied RareCollab to 890 patients from three cohorts, including 119 Undiagnosed Diseases Network probands with paired DNA and RNA data, constituting a large systematic benchmark for multimodal rare disease diagnosis under paired genomic and transcriptomic evaluation. In this real-world multimodal benchmark, RareCollab prioritized 94% of diagnostic genes within the top 10. Across recall thresholds from top 1 to top 10, it consistently outperformed proprietary phenotype-driven LLM baselines including Claude Sonnet 4.6 and GPT-5-mini by more than 25% on average and surpassed established state-of-the-art variant prioritization methods by 11%-24%. RareCollab also reshapes the diagnostic contribution of RNA evidence, which contributes to prioritization of the diagnostic gene in 35% of cases (42/119). Together, these results establish RareCollab as a scalable and interpretable framework for multimodal rare disease diagnosis.

## Introduction

Millions of children worldwide are affected by severe genetic diseases, many of which are Mendelian disorders[1–3]. For affected patients and families, obtaining a molecular diagnosis often involves a prolonged "diagnostic odyssey" that can extend for years[4]. Even after sequencing data are generated, pinpointing the disease-causal variant remains a major bottleneck because analyzing and interpreting diagnostic evidence is labor-intensive and time-consuming[5].

This bottleneck is being amplified by the growing diversity of diagnostic evidence. Rare disease diagnosis is increasingly informed by multiple complementary sources of evidence beyond phenotype comparison alone[6–7]. These include curated variant-level knowledge, computational predictions of variant pathogenicity, and other clinical or molecular data[8,9]. Among these modalities, transcriptome sequencing (RNA-seq) has emerged as a particularly informative diagnostic assay because it can reveal aberrant splicing and altered gene expression, thereby improving diagnostic yield beyond DNA-based analysis alone[10,11]. However, these evidence types differ fundamentally in format and distribution, and often include a mixture of structured features and unstructured textual information. Integrating them into a shared evaluative space requires complex logic to reconcile heterogeneous signals across modalities. In current practice, this step relies heavily on manual review[12,13] and manually curated rules[14] or features[15] that attempt to translate diverse evidence types into a shared evaluative space. This process is labor-intensive, difficult to scale and adds substantial burden to diagnostic interpretation. Therefore, a central challenge in rare disease diagnosis is the absence of computational approaches for systematically evaluating multimodal evidence within a common interpretive framework.

Recent studies have shown that large language models (LLMs) can support rare disease diagnosis, including phenotype-driven prompt-based ranking[16–19] and, more recently, agentic workflows that organize diagnostic reasoning[20]. However, many of these approaches still face the same bottleneck: integrating heterogeneous evidence within a shared evaluative framework. Most current LLM-based approaches remain centered on phenotypic description, with other evidence types incorporated only weakly or inconsistently into final prioritization. In addition, strong performance on relatively simplified benchmarking tasks does not necessarily translate to the complexity of real-world diagnostic evaluation[19]. Prompt-based ranking is particularly limited in this setting: in our previous work, such approaches produced outputs that were difficult to standardize and sensitive to candidate order[16], making them poorly suited to the structured synthesis of multimodal evidence required for clinical interpretation. Together, these limitations point to the need for LLM-based diagnostic frameworks that enable reliable and structured multimodal reasoning rather than single-prompt candidate ranking.

To address this challenge, we developed RareCollab, an LLM-powered multimodal diagnostic framework for rare diseases. Rather than relying on phenotype-driven prompting or asking a single model to jointly interpret all available evidence at once, RareCollab decomposes diagnosis into domain-specific reasoning over distinct lines of evidence, followed by structured cross-modal integration. This design supports variant-level synthesis of genomic, phenotypic, transcriptomic and curated biomedical evidence within a unified diagnostic workflow and enables variant prioritization in a shared evaluative space. In this study, we apply RareCollab to 890 patients from three cohorts, including 390 solved cases and 500 unsolved cases. We benchmark its performance against one-stop multimodal LLM baselines, phenotype-driven baselines using state-of-the-art proprietary LLMs, and established variant-prioritization frameworks. We further examine the contribution of transcriptomic evidence to diagnostic prioritization and evaluate the cross-cohort generalizability of RareCollab across distinct diagnostic settings.

## Results

### Overview of the RareCollab framework

RareCollab is a multimodal diagnostic framework that transforms fragmented diagnostic signals into a unified and interpretable prioritization of candidate variants through three sequential steps: candidate nomination, LLM Lab reasoning and contextual scoring. RareCollab defines the diagnostic search space through two complementary nomination routes derived from DNA and RNA evidence. DNA candidates are nominated by a Mixture-of-Experts (MoE) DNA Prioritization Engine, with nomination further adjusted by variant-calling confidence and concordance between observed zygosity and expected mode of inheritance. RNA candidates are nominated from abnormal splicing and abnormal expression detected by external bioinformatics tools[21, 22]. For each nominated candidate, RareCollab then applies four domain-specific LLM Labs to evaluate distinct lines of evidence: The RNA Lab assesses transcriptomic support from aberrant expression, aberrant splicing and allele-specific expression (ASE), while also evaluating their consistency with gene-level mechanisms, including constraint and dosage sensitivity; the Phenotype Lab evaluates phenotypic concordance using Human Phenotype Ontology (HPO)[23], curated disease descriptions from the OMIM database[24], and literature-derived knowledge; The Database Lab examines prior clinical and biological evidence from ClinVar[25] submission records; and the In-Silico Lab evaluates computational evidence for molecular impact, including predicted consequences of amino acid alteration and evolutionary conservation. These calibrated assessments are finally integrated through contextual scoring, enabling multimodal evidence to be compared within a shared evaluative space and translated into a clinically interpretable ordering of candidate variants.

### UDN cohort for multimodal diagnostic benchmarking

We next evaluated RareCollab in the Undiagnosed Diseases Network (UDN)[26], a setting well suited for testing multimodal diagnostic frameworks because it combines deep phenotyping with clinically resolved genomic and transcriptomic data (**Sup Fig. 1a, b**). From the 202 diagnosed probands with paired RNA-seq and DNA-seq, 119 met two uniform inclusion criteria for benchmarking (**Fig. 1c**): (1) the causative variant was reported in the UDN Gateway with a versioned transcript and cDNA change that allowed unambiguous resolution of genomic coordinates; and (2) the variant was successfully called by DeepVariant[27]-based variant calling. This design ensured that all included cases had consistently defined ground-truth variants for method comparison. The cohort showed a balanced sex distribution (68 males and 51 females) and was enriched for childhood- and adolescent-onset disorders. Genomic data were predominantly generated by whole-genome sequencing (WGS; n=95), with the remainder by whole-exome sequencing (WES; n=24). Among the 119 probands, 115 had blood RNA-seq and 79 had matched fibroblast RNA-seq, enabling cross-tissue assessment of transcriptomic abnormalities. Cases were recruited across multiple UDN clinical sites and spanned a broad range of specialties and phenotypic presentations (**Sup Fig. 1c**). Together, these features establish a clinically heterogeneous and molecularly rich benchmark for evaluating RareCollab under the complexity of real-world Mendelian disease diagnosis.

**Evaluation settings and comparator definitions**

Multimodal rare disease diagnosis poses two practical constraints for LLM-based evaluation: information security and computational tractability. Detailed variant-level evidence cannot be provided to proprietary LLMs under patient privacy constraints, so we standardized LLM benchmarking at the gene level. Because RareCollab itself performs prioritization at the variant level, gene-level benchmarking was derived by re-ranking candidate genes according to the highest-ranked variant within each gene. In addition, the full candidate set in a patient is typically too large to support exhaustive prompt-based ranking within current context-window limits. We therefore confined the LLM benchmark to the candidate search space nominated by RareCollab from DNA and RNA evidence. In the UDN benchmark cohort, this nominated space retained 97% of diagnostic genes while still containing approximately 40–61 candidate genes per case on average (mean, 43.8), indicating that it preserved most true diagnostic genes without collapsing the task into an overly simplified ranking problem (**Sup Fig. 2a–c**). To balance computational efficiency with patient privacy protection, RareCollab used a locally deployed GPT-oss-20B model as the foundation for all four LLM Labs, thereby avoiding the need to expose detailed candidate-level evidence to proprietary systems.

Within this matched benchmark setting, we defined two LLM baseline strategies: (1) a phenotype-driven baseline, a common setup in LLM-based rare disease diagnosis, in which candidate genes were ranked from patient phenotypes together with model-internal knowledge of gene function and disease association. This baseline therefore reflects how well an LLM can perform when it is allowed to rely entirely on its pretrained biomedical knowledge. (2) a one-stop multimodal baseline, in which all available multimodal evidence was provided directly to a single LLM without domain-specific decomposition. Because multimodal evidence substantially increased prompt length as well as time and computational cost, this one-stop multimodal baseline was evaluated only with GPT-oss-20B. To keep each query within a manageable context length, it was implemented by scoring each gene independently and then ranking genes by score, rather than ranking the full candidate set in a single prompt.

Within the overall evaluation setting, we evaluated RareCollab in three modes: (1) In the default mode, we sought to approximate real clinical application by using the full RareCollab framework with all available evidence channels enabled, including DNA, RNA, phenotype, database and in-silico evidence. In this setting, RareCollab was benchmarked against both LLM baselines as well as established variant prioritization methods. AI-MARRVEL (v1) was run with its default module, and Exomiser (v14) was run in its default configuration under recommended parameter settings[28] with the ClinVar whitelist enabled and REVEL[29], MVP[30], AlphaMissense[31] and SpliceAI[32] used as pathogenicity scores. (2) In the database-restricted mode, we modeled a more challenging benchmark scenario in which direct prior variant-level documentation was unavailable or limited. For RareCollab, this mode removed both the Database Lab and variant-level documentation-derived features from the DNA Prioritization Engine. For AI-MARRVEL, the Novel Disease Gene (NDG) module was used. For Exomiser, the ClinVar whitelist was disabled. (3) In the DNA-only mode, we modeled the more common diagnostic setting of WES or WGS without paired RNA-seq. In this mode, RNA evidence was removed from both candidate nomination and downstream reasoning.

## Diagnostic performance of RareCollab over LLM baselines

Under these matched conditions, RareCollab substantially outperformed both baseline strategies. Using the same GPT-oss-20B backbone, RareCollab achieved 62%, 90% and 94% top-1, top-5 and top-10 recall of diagnostic genes, respectively, compared with 12%, 48% and 61% for the phenotype-driven GPT-oss-20B baseline and 14%, 35% and 49% for the one-stop multimodal GPT-oss-20B baseline (**Fig. 2a**). Scoring with one-stop multimodal did not materially improve over phenotype-driven prioritization, indicating that simply providing more heterogeneous evidence to one LLM is insufficient for effective diagnostic ranking. RareCollab also outperformed the strongest phenotype-driven proprietary LLM baseline, Claude Sonnet 4.6, which reached 33%, 60% and 79% at top 1, top 5 and top 10, respectively (**Fig. 2b**). Together, these results indicate that RareCollab improves performance by decomposing multimodal diagnostic evidence into domain-specific reasoning followed by structured integration.

Reproducibility is also an important consideration in LLM-based rare disease diagnosis, where stochastic generation can introduce repeat-to-repeat variability in candidate ranking. Across five independent runs of RareCollab, top 1 through top 10 recall showed no significant difference by Cochran's Q test (**Sup Table 6, 7**), indicating stable aggregate performance across repeats. To quantify repeat-level variability at the gene level, we defined instability as $2p(1-p)$, where $p$ is the proportion of the five repeats in which a diagnostic gene was prioritized within top $K$ ($K$ = 1, 2, …, 10). This metric can be interpreted as the probability that two randomly selected repeats disagree on whether a diagnostic gene is prioritized within top $K$. Under this definition, RareCollab showed a marked reduction in instability as the review range expanded, decreasing from 9.3% at top 1 to 3.5% at top 3, and remaining low thereafter (**Sup Table 8**). By contrast, baseline models showed more variable instability profiles. In paired permutation tests, RareCollab showed significantly lower instability than every baseline model from top 3 onward (**Sup Table 9**). Notably, if top 10 is taken as a practical range for downstream manual review, RareCollab exhibited a mean instability of only 2.0%, indicating highly reproducible gene prioritization across repeated runs.

## Convergence of multimodal evidence in diagnostic prioritization

To pinpoint the advantage of integrating multimodal evidence, we analyzed how diagnostic variants were supported across modalities. Diagnostic genes/variants frequently showed concurrent strong support from multiple components of the framework, including DNA prioritization, RNA, phenotype, database and in silico evidence (**Fig. 2c** and **Sup 4a**). This pattern suggests that RareCollab does not depend on any single modality. Consistently, in ablation analyses of the contextual scoring stage, removing any single modality reduced top-10 recall by no more than 6% (**Sup Fig 5a, b**). This multimodal convergence was also associated with a marked narrowing of the diagnostic search space. As the number of strong evidence sources increased, the candidate space contracted sharply. Cases with only one strong evidence source contained 55 candidate variants (16 candidate genes) on average. With two strong evidence sources, the candidate space decreased to 10 variants and 7 genes on average; with three, it was reduced further to just 2 variants and 2 genes (**Fig. 2d** and **Sup Fig. 4c**). Thus, the evidence structure is potential to provide an interpretable way to reduce review burden at different confidence levels. Crucially, RareCollab coupled search-space reduction with effective prioritization of diagnostic genes and variants (**Fig. 2e** and **Sup Fig. 4d**): 85% of diagnostic genes or variants were supported by at least two strong evidence sources. This pattern was also reflected in ranking performance (**Fig. 2f** and **Sup Fig. 4e**): top-1 recall increased by 17%–25% with each additional strong evidence source. Once at least three strong evidence sources were present, top-5 recall reached 96% or higher. Together, these results show that RareCollab concentrates multimodal support onto true disease-causal candidates in an organized and interpretable framework that can facilitate efficient and accurate prioritization of clinically important variants.

### Benchmarking against state-of-the-art variant prioritization methods

We next benchmarked RareCollab against current state-of-the-art diagnostic methods under both default and database-restricted settings. Under default benchmarking, RareCollab consistently outperformed both Exomiser and AI-MARRVEL (**Fig. 3a** and **Sup Fig. 6a**). Compared with AI-MARRVEL, RareCollab improved top-1, top-5 and top-10 recall by 9%-11%; compared with Exomiser, the gains were 20%-27%. This strong performance was maintained across cases spanning different UDN confidence levels, from *Certain* cases with well-established disease mechanisms to *Highly Likely* cases that were more diagnostically challenging (**Sup Fig. 6b, c**). In the database-restricted setting, RareCollab remained the strongest-performing method, with 39% top-1 and 79% top-10 recall (**Fig. 3b**). These results show that RareCollab outperforms established variant prioritization methods widely used in current diagnostic practice, highlighting its potential for real-world rare disease diagnosis.

### Interpretable multimodal reasoning in a representative diagnostic case

A representative example of RareCollab's multimodal reasoning is a UDN case with compound-heterozygous pathogenic variants in NBAS[33], consisting of a nonsense allele (c.4753C>T) and a deep intronic splice-altering allele (c.2423+403G>C) (**Fig. 3c**). This case is diagnostically challenging because it requires resolving both compound heterozygosity and a non-coding pathogenic variant. Even without database evidence, RareCollab ranked the nonsense variant first and the splice-altering variant second, whereas Exomiser ranked the gene 19th and AI-MARRVEL ranked the nonsense variant 17th while failing to prioritize the splice variant within the top 100. This advantage arose from convergent support across multiple evidence domains. The RNA Lab identified ASE, potential exon skipping and gene-level transcriptomic abnormalities consistent with expected mechanisms. The Phenotype Lab found strong concordance between NBAS and the patient phenotype across multiple phenotypic axes, including growth, skeletal, ocular and neurodevelopmental features. The In-Silico Lab provided strong support for the nonsense allele through deleteriousness and conservation signals. This case illustrates how RareCollab prioritizes causal variants in a challenging real-world patient case while preserving tractable reasonings.

## Contribution of transcriptomic evidence to multimodal diagnostic prioritization

The findings from the case example also raise a broader question about the contribution of transcriptomic evidence to computational variant prioritization, which has not been systematically quantified. Because diagnostic variants with RNA-level abnormalities are often research variants that are more likely to have prior variant-level documentations, directly assessing RNA contribution in the default setting may be confounded by database support. To mitigate this bias, we systematically evaluated RNA contribution under the database-restricted setting. In this setting, excluding RNA reduced top-1, top-5 and top-10 recall by 6%–11% (**Fig. 4a**). These results show that RNA is a substantial contributor to diagnostic performance when variant documentation is not available.

We next found that RNA data contributes to diagnosis through two distinct paths. We defined RNA-driven events as diagnostic variants nominated only through RNA evidence, and RNA-supported events as diagnostic variants already nominated from DNA but moved to a higher rank with RNA evidence. Across the cohort, we identified 16 RNA-driven and 26 RNA-supported diagnostic variants (**Fig. 4b**). The underlying RNA signals differed between these groups (**Fig. 4c**): expression abnormalities were more prominent among RNA-driven events, whereas splicing abnormalities were more frequent among RNA-supported events. The resulting rank shifts were substantial (**Fig. 4d**). Without RNA, none of the RNA-driven variants ranked within the top 20; with RNA, 11 (69%) were promoted into the top 5 and 14 (88%) into the top 10. RNA also sharpened prioritization after DNA nomination: among the 26 RNA-supported variants, 11 (42%) improved by at least five ranking positions. Notably, RNA contribution was usually not isolated. Instead, RNA-contributed diagnostic genes were most often supported in combination with phenotype and DNA evidence, rather than by RNA alone (**Fig. 4e**). This pattern suggests that RareCollab does not simply add RNA as an independent diagnostic signal, but reshapes the contribution of transcriptomic evidence through its interaction with other modalities.

## Representative cases of RNA-driven and RNA-supported prioritization

Representative UDN cases illustrate the two distinct ways in which RNA contributes to RareCollab prioritization. The HADHB case[34] is shown as an RNA-driven example: the diagnostic variants were not effectively captured by DNA-based prioritization alone, but strong transcriptomic abnormalities enabled RNA-based nomination and promoted them to the top ranks after multimodal integration (**Fig. 4f**). By contrast, the FLCN case illustrates RNA-supported reprioritization: the variant had already been nominated from DNA, but RNA revealed decreased expression consistent with its expected mechanism, which strengthened the overall evidence profile and further improved its rank (**Fig. 4g**). Additional RNA-driven examples are shown for an MECR case[35] and a second HADHB case (**Sup Fig. 7a, b**). In dual-modality orthogonal analyses, these diagnostic variants consistently received stronger support from RNA together with one or more additional evidence domains than non-diagnostic variants from the same case (**Sup Fig. 8 a-d**), further showing that RNA contribution in RareCollab is typically not isolated but reinforced through cross-modal agreement. Together, these examples show that RareCollab treats RNA as an interpretable source of diagnostic evidence that can both uncover missed pathogenic variants and materially reshape candidate prioritization.

## Cross-cohort generalizability in DNA-only diagnostic settings

Because most real-world diagnostic workflows rely on WES or WGS without paired RNA data, we evaluated RareCollab in the DNA-only setting across cohorts to assess generalizability. We analyzed three cohorts representing distinct diagnostic contexts: the 119-proband UDN benchmark, a deeply phenotyped and diagnostically challenging cohort; a Baylor Genetics (BG) cohort of 58 cases reflecting routine clinical laboratory practice; and the publicly available Deciphering Developmental Disorders (DDD) cohort of 213 cases, a widely used benchmark for developmental-disorder gene discovery. These cohorts differed in phenotype density and overall data distribution. BIOLORD[36] embeddings of patient HPO profiles suggested differences in phenotype representation across cohorts, and HPO term counts indicated differences in phenotype density (**Fig. 5a, b**). UDN cases showed greater phenotype complexity and a larger candidate space, but in all three cohorts the number of candidate variants decreased as more strong evidence sources were required, and diagnostic genes were typically supported by multiple strong evidence sources rather than isolated signals (**Fig. 5c–f**).

Across these heterogeneous cohorts, RareCollab maintained strong diagnostic performance, with top-10 recovery exceeding 90% in all three settings (**Fig. 5g**). To quantify cross-cohort stability, we calculated, at each rank threshold, the gap between the highest and lowest recall observed across the three cohorts and then averaged these gaps across top 1–10. RareCollab showed a mean cross-cohort gap of only 8%, substantially smaller than that of AI-MARRVEL (17%) or Exomiser (18%) (**Fig. 5g**). These results indicate that RareCollab not only achieves strong performance across distinct diagnostic settings but also generalizes more stably than existing methods when transferred across independent cohorts.

## Discussion

RareCollab addresses a central limitation of rare disease diagnosis: diagnostic evidence is often distributed across DNA, RNA, phenotype, database knowledge, and in-silico prediction, yet these signals are rarely evaluated within a shared framework. By bringing these heterogeneous signals into an interpretable and portable diagnostic framework, RareCollab outperformed both LLM baselines and widely used variant prioritization methods, with robust performance across cohorts and diagnostic settings.

Large language models have attracted growing interest in rare disease diagnosis by raising the possibility that fragmented clinical and biomedical knowledge can be transformed into scalable decision support with real-world clinical impact[37]. This momentum has increasingly driven efforts to couple LLMs with clinically grounded knowledge and decision structure to achieve more interpretable and robust diagnostic strategies[20,38]. However, these efforts have remained centered largely on gene–phenotype matching. RareCollab builds on this direction by extending LLM-powered diagnosis beyond phenotype matching alone to the structured evaluation and integration of multiple diagnostic evidence modalities within a single workflow.

Several limitations should be noted. First, paired DNA–RNA benchmark remains modest in size, reflecting the scarcity of clinically resolved multimodal rare disease cohorts. However, as RNA-seq becomes more accessible[39] and increasingly incorporated into rare disease evaluation, this setting is likely to become more common and clinically impactful. Second, the current implementation of RareCollab does not yet encompass the full range of diagnostic modalities that are becoming increasingly relevant in rare disease diagnosis, including copy-number variation[40], epigenetic profiling[41], and non-coding RNA[42]. As rare disease diagnosis continues to incorporate a broader range of evidence beyond DNA evidence and phenotypic evidence, the value of integrating these diverse modalities within a shared interpretive framework is showing increasing potential. Third, error analysis showed that, if top 10 is taken as a practical threshold for expert review, nomination failure accounted for half of failed cases in the default setting and nearly two-thirds under database restriction (**Sup Fig. 9a, b**). Thus, further improving the step that narrows thousands of variants to a short list may yield substantial additional gains.

Finally, applying RareCollab to 500 unsolved UDN cases. On average, each unsolved case contained two variants supported by at least three strong evidence sources (**Sup Fig 10a**). Even when the DNA Prioritization Engine did not rank a variant within the top 10, RareCollab still identified two variants on average per case with support from more than two strong evidence sources, with RNA and phenotype representing the most frequent contributing categories (**Sup Fig 10b**). These findings suggest that RareCollab may provide a practical framework for systematic review of new candidate variants beyond conventional DNA-driven pipelines.

Together, these results support RareCollab as an interpretable multimodal diagnostic framework that improves prioritization of clinically important variants, clarifies the contribution of transcriptomic evidence, and provides a foundation for more systematic evaluation of unresolved rare disease cases.

## Methods

### Data sources and preprocessing

DNA sequencing and RNA-seq data were obtained from the Undiagnosed Diseases Network (UDN) study deposited in dbGaP under accession phs001232.v1.p1. Phenotype annotations and clinical reports were retrieved from the UDN Gateway. Together, these resources provided the molecular and clinical inputs for candidate-level interpretation.

For DNA preprocessing, raw whole-exome and whole-genome FASTQ files were aligned to GRCh38/hg38 using bwa-mem2 (v2.2.1). Alignments were processed with samtools (v1.22.1), including name sorting, mate fixing, coordinate sorting, duplicate marking and BAM indexing. Small variants were then called from the processed BAM files using DeepVariant (v1.9.0) and exported in VCF format. These VCF files were subsequently processed through the AI-MARRVEL v1 preprocessing pipeline, which annotated each candidate variant with 101 curated features for downstream DNA-based prioritization.

For RNA preprocessing, raw RNA-seq FASTQ files were processed with the nf-core/rnaseq pipeline implemented in Nextflow (v3.21.0) to generate analysis-ready BAM files aligned to GRCh38/hg38. RNA-seq alignments were then analyzed using the DROP pipeline (development version) as the core framework for RNA outlier detection. Within DROP, OUTRIDER was used to detect aberrant expression and FRASER2 to identify aberrant splicing events. Allele-specific expression and mono-allelic expression analyses were performed using GATK (v4.6.2.0)[43], including GenotypeGVCFs and ASEReadCounter. This workflow enabled unified detection of multiple classes of transcriptomic abnormalities from the same RNA-seq dataset.

### LLM reproducibility

To assess reproducibility, all LLM-based methods were evaluated across five independent repeats on the UDN default benchmarking setting. For phenotype-driven LLM baselines, each repeat used the same prompt template, but the order of candidate genes was shuffled to reduce the influence of positional bias inherent to ranking tasks. Because all phenotype-driven baselines were evaluated with the same prompt set, their results remained directly comparable. For the one-stop multimodal-input LLM baseline and for RareCollab, each LLM query evaluated only a single candidate gene or variant rather than ranking an entire candidate list within one prompt. Therefore, all five repeats used the same prompt without shuffling.

For result reporting, LLM baseline performance was calculated by averaging the scores or ranks obtained across the five repeats and then re-ranking candidates accordingly. RareCollab showed no significant performance differences across five repeats (**Sup Table 4-9**), and no single repeat displayed a consistent advantage across evaluation metrics. Therefore, to reduce computational cost and runtime in cross-setting and cross-cohort analyses, the reported results for RareCollab were based on one representative repeat in the UDN default setting, whereas other benchmarking settings and cohorts were evaluated with a single run.

**RareCollab framework**

RareCollab is a multimodal variant-prioritization framework with three stages: candidate nomination, LLM reasoning, and contextual scoring. Candidate variants are nominated from complementary DNA-based and RNA-based evidence sources, evaluated by domain-specific LLM Labs, and then compared within each case through contextual scoring to generate an interpretable rank order for downstream review.

**Candidate nomination**

RareCollab defines the downstream search space through two complementary nomination routes: DNA-based prioritization and RNA-derived transcriptomic evidence. The final nominated set was defined as the union of these DNA- and RNA-derived candidates.

**DNA-based candidate nomination**

DNA-based nomination was performed using an MoE model trained on variant-level candidate annotations. Each candidate variant was represented by 101 curated features generated by the AI-MARRVEL preprocessing pipeline. These features were organized into four base evidence domains—Genetics, In-Silico, Database and Phenotype—together with an additional Overview domain defined as the union of all base-domain features. The Genetics domain included variant consequence, zygosity, inheritance compatibility, allele frequency, repeat context and gene-level constraint features. The In-Silico domain included pathogenicity predictors and conservation metrics. The Database domain captured prior gene- and variant-level support from curated databases. The Phenotype domain included phenotype-matching features such as Phrank[44] and related symptom-similarity scores. Detailed training procedures are provided in the **Supplementary Method**.

Each domain was processed by a dedicated expert implemented as a two-layer multilayer perceptron (MLP), which produced both a domain-specific embedding and a scalar domain logit. Domain embeddings were concatenated and passed to a fusion MLP to generate an overall diagnostic score for each candidate while retaining parallel domain-level scores for downstream use. Each expert used a hidden layer of 64 units followed by a 32-dimensional embedding layer. The fusion module used hidden dimensions of 64 and 32 with layer normalization and dropout.

After scoring, candidate variants were ranked within each case according to the overall score from the DNA Prioritization Engine as well as the parallel domain-specific scores from the Genetics, In-Silico and Database domains. A variant was nominated through the DNA route if it ranked within the top 20 candidates by the overall DNA score, or if it showed particularly strong support in one of the Genetics, In-Silico or Database domains by ranking within the top 10 of that domain while remaining within the top 100 candidates overall. To preserve potential compound-heterozygous configurations, if any selected variant within a gene was compatible with recessive mode-of-inheritance annotated on the OMIM database, all variants from the same gene were retained for downstream interpretation.

### RNA-based candidate nomination

RNA-based nomination served as a complementary nomination route based on transcriptomic outlier signals derived from the RNA preprocessing workflow. Candidates were nominated when they showed either strong evidence of junction-level aberrant splicing or gene-level aberrant expression.

For aberrant splicing, a candidate variant was retained if the FRASER2 adjusted *P* value was below 0.01 and the variant ranked among the 10 most significant splice-outlier variants in that case based on the corresponding splicing *P* value. For aberrant expression, a candidate gene was retained if the OUTRIDER adjusted *P* value was below 0.01 and the gene ranked among the 10 most significant expression outlier genes in that case based on the corresponding expression *P* value. RNA-derived candidates were then merged with DNA-nominated candidates for downstream interpretation.

### Variant-calling confidence

An auxiliary variant-calling confidence assessment was incorporated to identify candidates at increased risk of miscalling. This assessment combined two sources of concern: absence of alternate RNA support despite adequate reference coverage, and indel localization within highly repetitive sequence context.

The RNA-based signal was derived from allele-specific read counts in RNA-seq alignments. A candidate was flagged when the mean reference allele count exceeded 20 while the minimum alternate allele count was 0 across available RNA observations, consistent with possible under-calling of the alternate allele despite adequate reference support. In parallel, indel and duplication candidates were evaluated for local repeat context by scanning the 30-bp flanking sequence on both sides of the variant for short homopolymer or tandem-repeat patterns of length 1–6 bp. Candidates were flagged when the local sequence context contained more than 10 consecutive repeat units, indicating increased risk of alignment ambiguity. Only candidates without either source of concern were considered to pass the variant-calling confidence check.

### LLM Lab Reasoning

After nomination, each candidate was evaluated by four domain-specific modules: Phenotype Lab, Database Lab, In-Silico Lab and RNA Lab. Rather than asking a single model to rank all candidates directly, RareCollab decomposed interpretation into domain-restricted tasks, each operating on a predefined evidence slice and returning a structured categorical conclusion. To maintain computational tractability, LLM evaluation within each case was restricted to candidates meeting modality-specific entry criteria, thereby limiting the number of queries per Lab to a tractable range of approximately 40–50. Phenotype Lab was applied to all nominated genes without additional filtering. Database Lab was applied only to variants with ClinVar submission records. In-Silico Lab was applied only to candidates ranked within the top 100 of the In-Silico Domain of the DNA Prioritization Engine. RNA Lab was applied only to candidates with transcriptomic signals meeting at least one of the following criteria: a FRASER2 junction-level or gene-level $P$ value ≤ 0.1, a GATK ASE $P$ value ≤ 0.1, or an absolute OUTRIDER z score ≥ 1.5. These criteria were designed to avoid uninformative LLM queries and to reduce runtime and computational cost while preserving potentially meaningful diagnostic evidence. The original prompt templates and raw LLM outputs are provided in the **Supplementary Method**.

**Phenotype Lab**

The Phenotype Lab evaluated candidate–phenotype concordance through three complementary phenotype views: curated HPO-derived gene phenotypes, OMIM clinical feature descriptions and phenotype evidence extracted from the literature.

HPO-based phenotype matching. The HPO component compared the patient's HPO profile against gene-associated phenotype annotations downloaded from the HPO resource. In addition to phenotype terms, the model received observed-frequency annotations so that recurrent manifestations could be distinguished from less consistently reported features. Rather than relying on raw term overlap alone, the prompt instructed the model to organize reasoning by major phenotype axes or organ systems and to weigh both concordant features and important unexplained features. The output consisted of axis-structured reasoning together with one of four ordinal labels: Not Fit, Partial Fit, Good Fit or Stand-Alone Strong Evidence.

OMIM-based phenotype matching. The OMIM component evaluated concordance between the patient phenotype profile and gene-associated disease descriptions retrieved through API-based querying of OMIM. For each candidate gene, the associated disease entries were used to extract the *Description* and *Clinical Features* fields as input evidence. This component performed concise text-level matching without axis-based decomposition and emphasized both specific overlaps and major missing features. In addition to the four concordance labels above, this component also allowed an Impossible label when the OMIM record explicitly indicated that a required or invariable feature was absent from the patient phenotype profile. Candidates assigned Impossible were excluded from subsequent contextual scoring.

Literature-based phenotype matching. The literature component was applied only to genes lacking associated HPO annotations. This component was implemented in two stages. First, gene-associated articles were retrieved from PubMed by querying the gene symbol together with a predefined set of rare disease-related keywords, including *variant*, *mutation*, *diagnosis*, *Mendelian* and *cause*. PMID and title were then passed to a Literature Curator LLM, which retained articles relevant to rare disease gene–phenotype associations and assigned each to one of five study categories: phenotype association, functional study, patient cohort, case report or review. Second, the title and abstract of retained articles were passed to a Literature Phenotype Reasoning LLM, which evaluated whether the reported phenotype evidence provided meaningful support for the candidate gene in the context of the patient's phenotype profile. This step returned a concise rationale grounded in the selected literature together with one of three labels: Not Fit, Partial Fit or Good Fit.

**Database Lab**

The Database Lab interpreted prior variant-level evidence from ClinVar submission records. For each candidate variant, ClinVar was queried to retrieve submission history, including submitted classifications and supporting record-level details, and these records were incorporated directly into the prompt as structured evidence. Based on this information, the model was instructed to assess whether the overall ClinVar evidence argued against pathogenicity, was uninformative or supported pathogenicity, while also inferring the zygosity pattern explicitly stated or implied by the submissions. The output therefore contained two structured components: a pathogenicity-support label (Against, Neutral, Supporting or Convincing) and a zygosity label (heterozygous, homozygous, compound heterozygous or no information).

**In-Silico Lab**

The In-Silico Lab evaluated whether computational predictors and conservation metrics, considered in isolation from phenotype, population, segregation and database evidence, supported a pathogenic interpretation. Inputs included deleteriousness scores and conservation features, including CADD[45], DANN[46], REVEL[29], M-CAP[47], PolyPhen-2[48], SIFT[49], MutationAssessor[50], FATHMM[51], GERP[52], phyloP[53], LRT[54]-based measures, and SpliceAI[32]. The prompt explicitly framed this module as a domain-restricted assessment of computational support only. Output was standardized to one of three labels: Weak, Moderate or Strong.

**RNA Lab**

The RNA Lab interpreted transcriptomic evidence at both the gene and variant levels using outputs from FRASER2, OUTRIDER and ASE analyses, together with gene-level dosage sensitivity, loss-of-function constraint and inheritance context. Specifically, the prompt included ClinGen dosage sensitivity, probability of loss-of-function intolerance (pLI) and loss-of-function observed/expected ratio to contextualize the expected transcriptomic consequences of disruption in each gene, together with inheritance-related information including known disease mode-of-inheritance and zygosity. When multiple RNA-seq samples from different tissues were available for the same patient, the most significant $P$ value across tissues was used as the RNA Lab input.

At the gene level, the prompt incorporated gene-level metrics from FRASER2 and OUTRIDER. At the variant level, it incorporated junction-level splicing statistics from FRASER2, gene-level expression metrics from OUTRIDER, and ASE metrics. Based on these inputs, the LLM was tasked with classifying candidates into RNA events and assigning an overall conclusion of No RNA Evidence, Weak RNA Evidence or Strong RNA Evidence.

Across all four labs, outputs were restricted to compact predefined conclusion spaces rather than unconstrained natural-language judgments. This design converted heterogeneous candidate-specific evidence into standardized intermediate representations for downstream within-case contextual scoring. For visualization purposes, “strong evidence” was defined as Good Fit or Stand-Alone Strong Evidence for the Phenotype Lab, Strong RNA Evidence for the RNA Lab, Convincing for the Database Lab, and Strong for the In-Silico Lab. These cutoffs had no effect on the prioritization results.

**Compound-heterozygous configuration**

Potential compound-heterozygous configurations were reconciled within each case and gene after nomination and domain-specific evidence interpretation. Pairing was restricted to candidates compatible with recessive inheritance. A partner variant was considered eligible if it ranked within the top 50 nominated variants by the DNA Prioritization Engine score or carried orthogonal supporting evidence from other domains, including a frameshift consequence annotated by the Variant Effect Predictor (VEP)[55], strong RNA evidence assigned by the RNA Lab or convincing database evidence assigned by the Database Lab. Candidate pairs with incompatible allele-specific RNA patterns were excluded. Supported pairs were subsequently represented as compound entities during contextual scoring.

### Contextual scoring and final ranking

Final prioritization was performed by contextual scoring of pooled nominated candidates. Rather than relying solely on upstream nomination scores, candidates within the selected nomination tiers were re-evaluated jointly within each case to account for the local diagnostic context. Candidates outside this pooled set were retained below the rescored pool with their original ordering.

Contextual comparison was performed at the entity level. Variants forming a potential compound-heterozygous configuration within the same case were grouped into a single compound entity, whereas all other candidates were treated as singleton entities. When a compound entity was compared with a singleton entity, entity-level evidence for the compound entity was represented by the stronger of its two constituent variants. When two compound entities were compared, entity-level evidence for each compound entity was summarized by the mean value across its constituent variants.

Within each case, pooled entities were compared using Copeland pairwise majority voting. Each pairwise comparison considered six primary evidence dimensions: DNA Prioritization Engine rank, rna support, phenotype support, database support, inheritance match and frameshift status. Besides core modalities of RareCollab, inheritance match was defined for all variants and set to 0 specifically for heterozygous variants in autosomal recessive genes when no supported compound-heterozygous configuration could be established; otherwise, it was set to 1. Frameshift status was derived from VEP annotation. The entity favored on a given dimension received one vote, and the entity with more votes was preferred. If the primary vote was tied, in-silico support was used as a tie-breaker; if the tie persisted, the entity with the better baseline DNA Prioritization Engine rank was preferred.

Final ranking was then derived from the Copeland score, defined as the number of pairwise wins minus the number of pairwise losses for each entity within the same case. Entities with higher Copeland scores were ranked higher, with residual ties resolved by the total number of wins, then losses, and finally the DNA Prioritization Engine rank. The resulting entity-level order was then mapped back to variant-level rankings, enabling multimodal evidence to be interpreted comparatively within each case.

## Data and Code Availability

The DNA and RNA sequencing data analyzed in this study are available through dbGaP. UDN Gateway metadata used in this study require authorization from the Undiagnosed Diseases Network (UDN) and can be obtained through the corresponding UDN access procedures. Data from the BG cohort are in-house and are therefore not publicly available. Data from the DDD cohort are publicly available and can be accessed through the corresponding data access procedures established by the Deciphering Developmental Disorders (DDD) study.

RareCollab is distributed as an installable software package. The source code is available at https://github.com/LiuzLab/RareCollab.

## Acknowledgments

This study was primarily supported by the Chao Endowment and the Huffington Foundation to Zhandong Liu. Shinya Yamamoto, Hu Chen and Zhandong Liu received support from the Chan Zuckerberg Initiative (#2023-332162, #2025-367725). Pengfei Liu was also supported by the National Human Genome Research Institute (R35HG011311). The Undiagnosed Diseases Network is supported by the National Institute of Neurological Disorders and Stroke of the National Institutes of Health (U2CNS132415, U01HG010218, U01HG007708). We thank Hyun-Hwan Jeong for facilitating access to the dbGaP data. The content is solely the responsibility of the authors and does not necessarily represent the official views of the National Institutes of Health.

## Author Contributions

Conceptualization: G.Q., H.C., Z.L.; Data Curation: G.Q., J.W., Z.S., M.R., D.B., J.C., K.S., M.W., S.B., J.B.; Formal Analysis: G.Q., J.W.; Funding acquisition: Z.L.; Investigation: G.Q., J.W., M.L.C., S.L., S.Y.; Methodology: G.Q.; Visualization: G.Q., J.W., S.P.; Writing – original draft: G.Q.; Writing – review & editing: G.Q., J.W., H.C.; Supervision: Z.L., H.C., P.L.

## Conflict of interest

All authors declare that they have no conflict of interest.

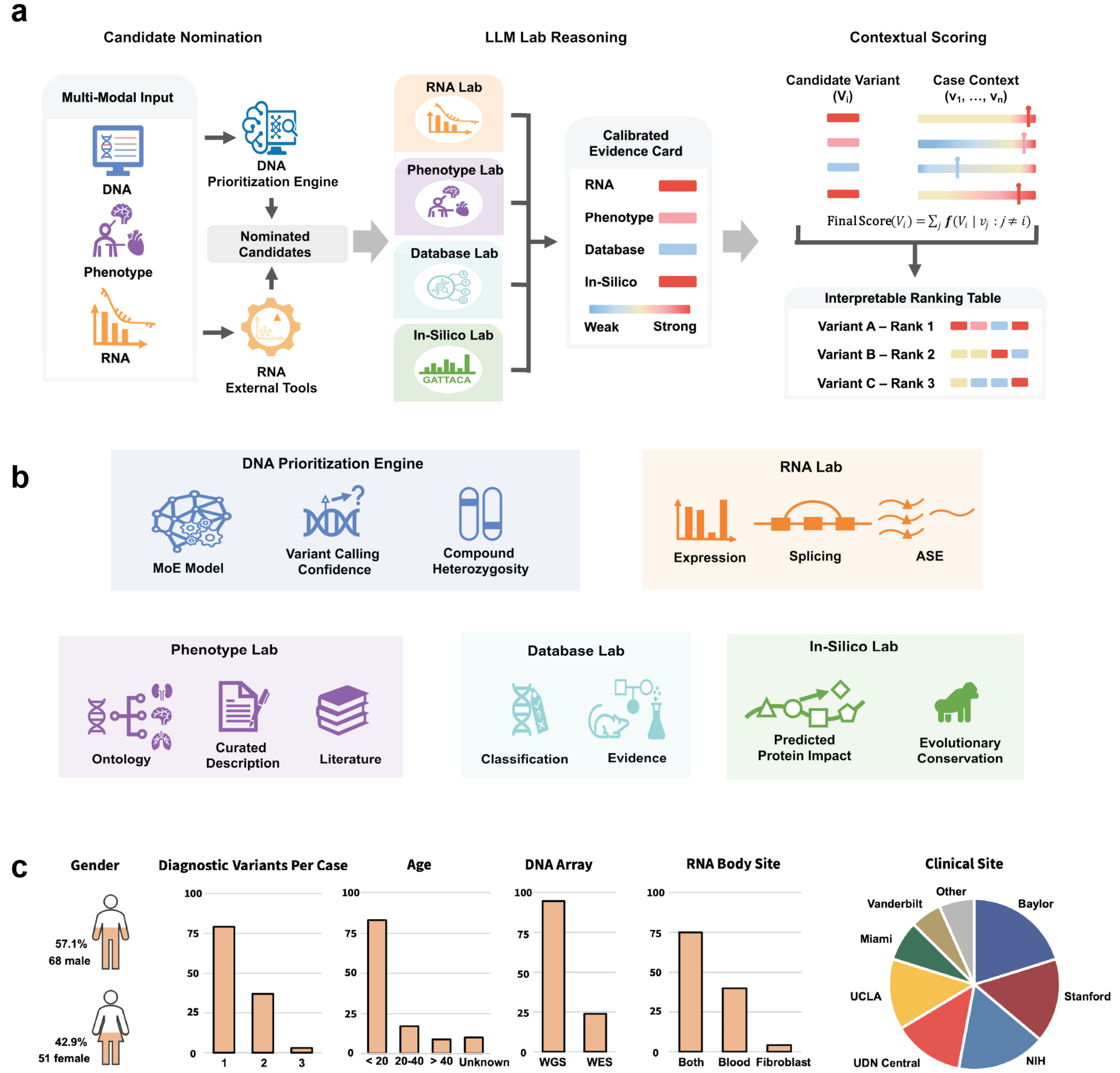

**Figure 1 | RareCollab framework and benchmark cohort.**
**a,** Overview of RareCollab. Case data include DNA, phenotype, and RNA. Candidates are first nominated by the DNA Prioritization Engine and RNA-based external tools. The nominated candidates are then evaluated by four domain-specific LLM Labs: RNA Lab, Phenotype Lab, Database Lab, and In-Silico Lab. These modules convert heterogeneous evidence into calibrated conclusions. Final prioritization is performed by contextual scoring, which compares candidates within the same case and generates an interpretable ranking table.
**b,** Inputs used by each module. The DNA Prioritization Engine incorporates a mixture-of-experts model, variant-calling confidence, and compound-heterozygous pairing. The RNA Lab evaluates aberrant expression, aberrant splicing, and allele-specific expression. The Phenotype Lab integrates ontology-based information, curated disease descriptions, and literature. The Database Lab captures prior classifications and supporting evidence from ClinVar submission records. The In-Silico Lab evaluates predicted protein impact and evolutionary conservation.
**c,** Overview of the benchmark cohort, including sex, number of diagnostic variants per case, age, DNA sequencing modality, RNA tissue source, and contributing clinical sites.

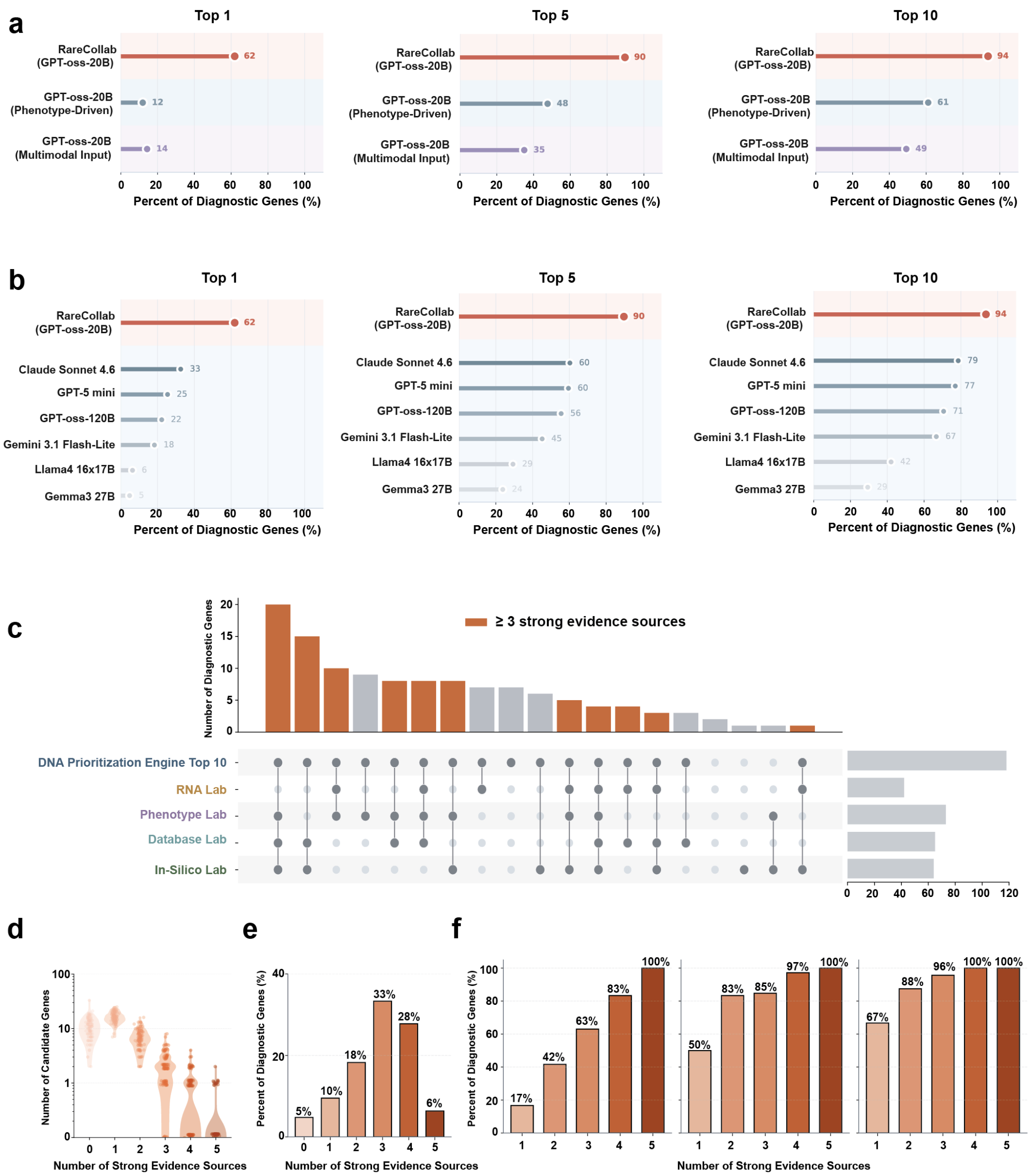

**Figure 2 | Benchmarking performance and evidence structure of RareCollab.**

**a,** Comparison of RareCollab with two GPT-oss-20B baselines, including a phenotype-driven baseline and a one-stop multimodal-input baseline, at top 1, top 5, and top 10.

**b,** Comparison of RareCollab with phenotype-driven baselines built on multiple phenotype-driven LLM baselines at top 1, top 5, and top 10.

**c,** UpSet-style summary of strong evidence-source combinations supporting diagnostic genes. Groups with at least 3 strong evidence sources are highlighted. Combination counts are shown above the matrix and per-domain totals are shown at right.

**d,** Number of candidate genes as a function of the number of strong evidence sources supporting the diagnostic gene.

**e,** Distribution of diagnostic genes by the number of strong evidence sources.

**f,** Recovery of diagnostic genes at top 1, top 5, and top 10, stratified by the number of strong evidence sources.

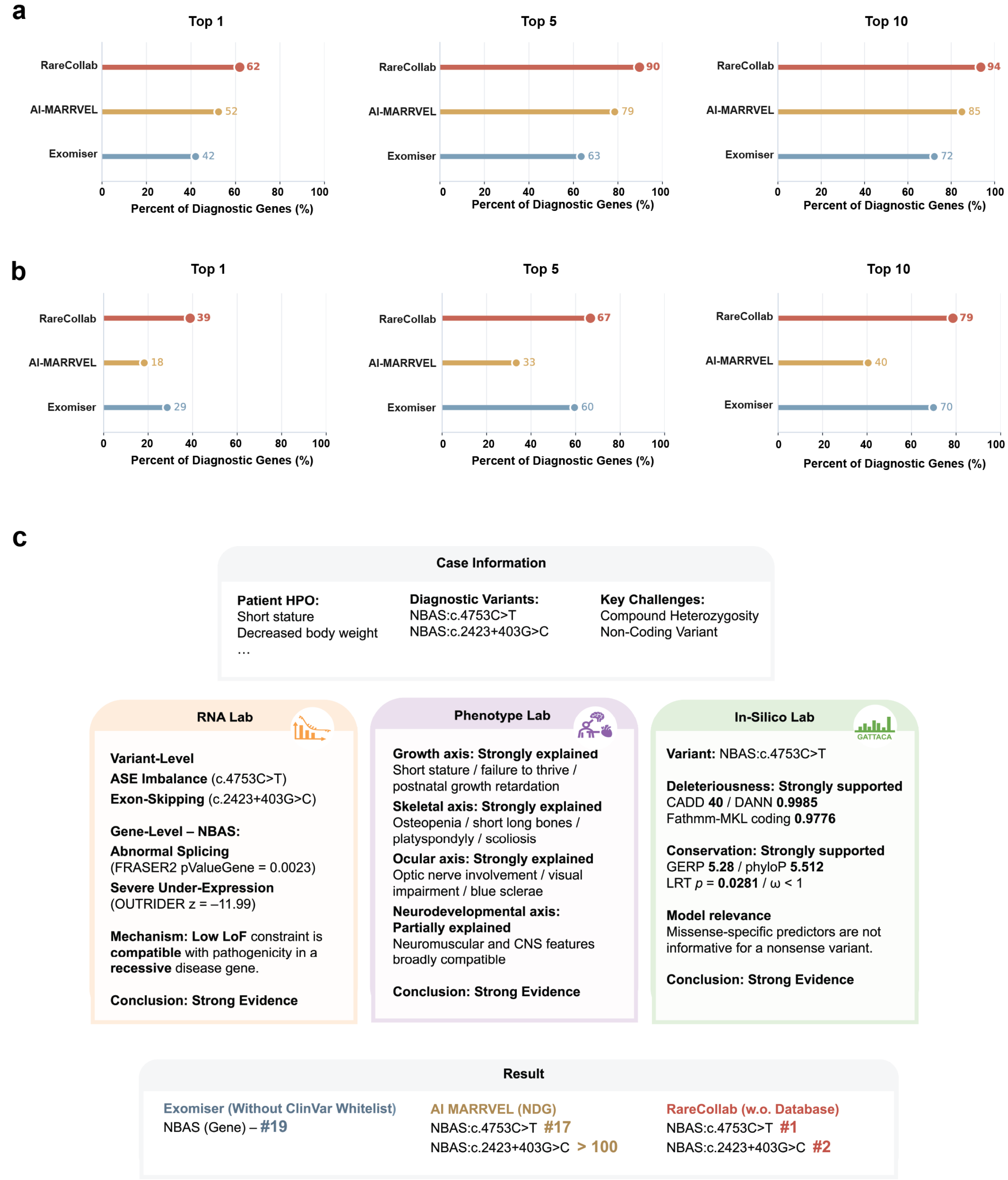

**Figure 3 | Benchmarking of RareCollab against current diagnostic methods under default and database-restricted settings.**

**a,** Standard benchmarking of RareCollab, AI-MARRVEL, and Exomiser at top 1, top 5, and top 10.

**b,** Database-restricted benchmarking of RareCollab, AI-MARRVEL, and Exomiser at top 1, top 5, and top 10.

**c,** Representative database-restricted compound-heterozygous case illustrating multimodal reasoning in RareCollab and rankings from RareCollab, Exomiser, and AI-MARRVEL.

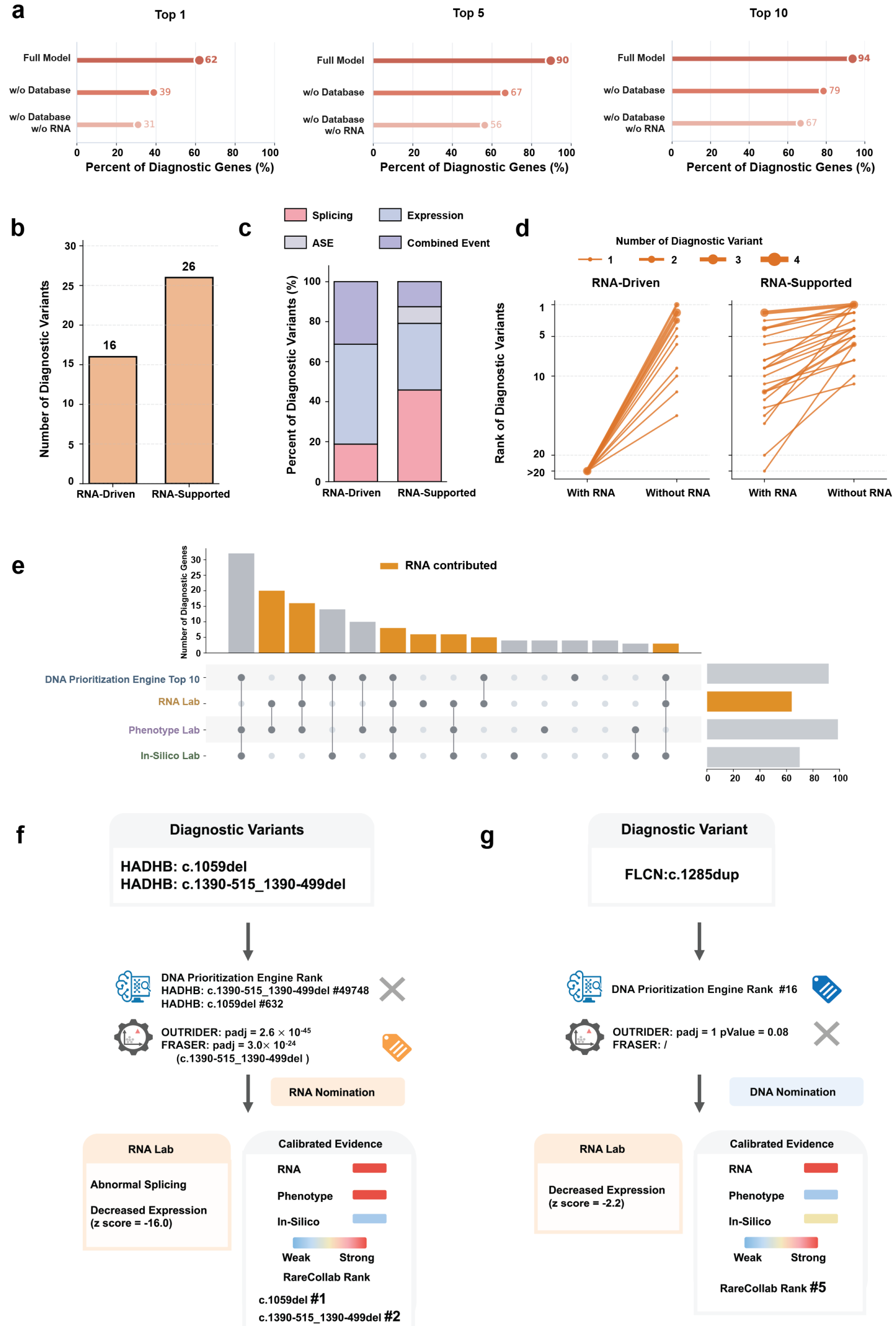

**Figure 4 | RNA contribution to diagnostic prioritization in RareCollab.**
**a,** Recall at top 1, top 5, and top 10 for the full model, database-restricted model, and the model without both database and RNA evidence.
**b,** Number of diagnostic variants classified as RNA-Driven or RNA-Supported.
**c,** Distribution of LLM-identified RNA event types in the RNA-Driven and RNA-Supported groups.
**d,** Diagnostic variant ranks with and without RNA evidence in the RNA-Driven and RNA Supported groups.
**e,** UpSet-style summary of strong evidence-source combinations in the database-restricted setting. Groups with RNA contribution are highlighted. Combination counts are shown above the matrix and per-domain totals are shown at right.
**f,** Representative RNA-Driven case illustrating RNA-based nomination and subsequent prioritization of diagnostic HADHB variants.
**g,** Representative RNA-Supported case illustrating DNA-based nomination with additional support from RNA evidence for a diagnostic FLCN variant.

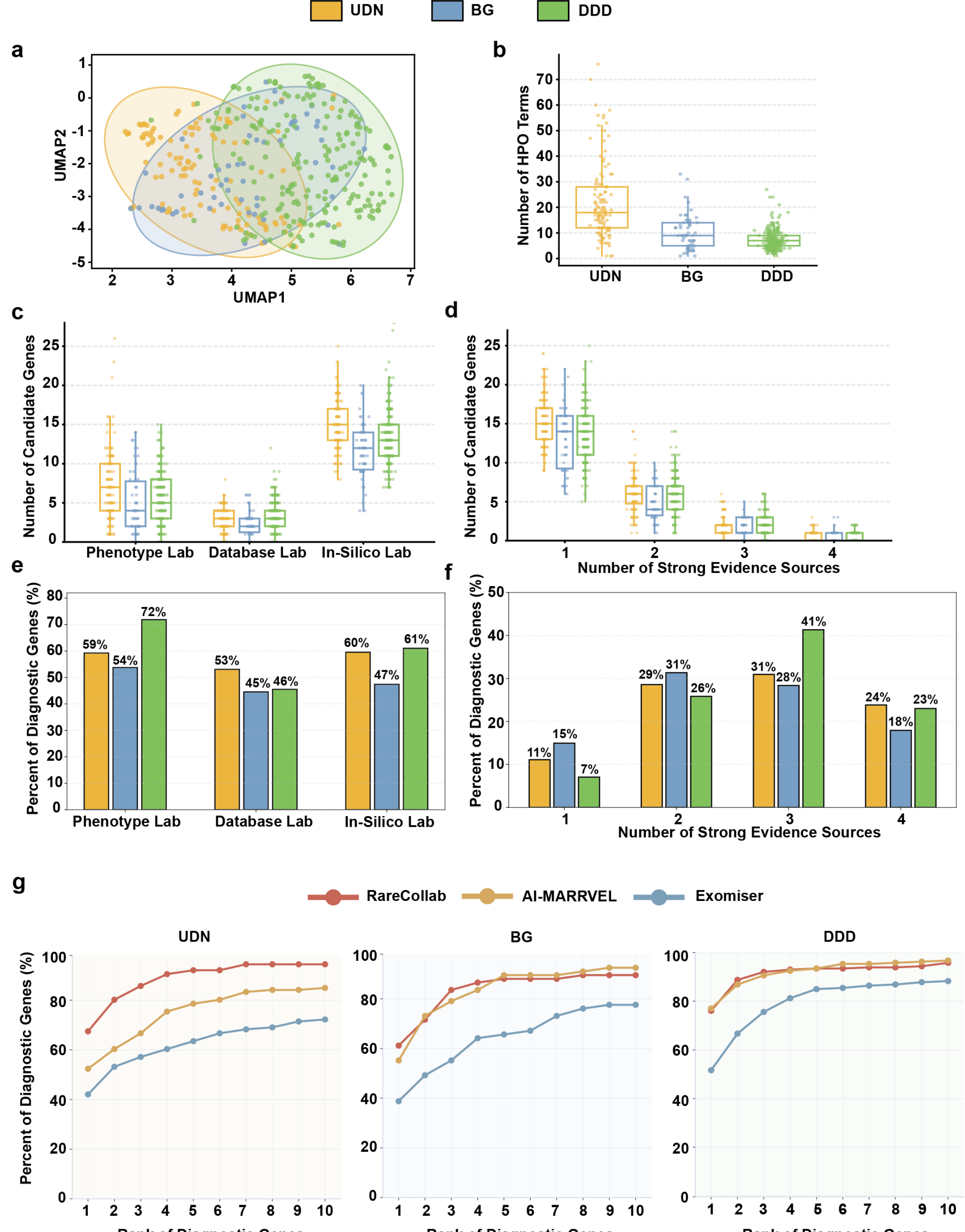

**Figure 5 | Cross-cohort evidence structure and generalizability of RareCollab.**

**a,** UMAP projection of phenotype embeddings from the UDN, BG, and DDD cohorts.

**b,** Distribution of phenotype complexity across cohorts, measured by the number of HPO terms per case.

**c,** Number of candidate genes supported by the Phenotype, Database, and In-Silico Labs in each cohort.

**d,** Number of candidate genes as a function of the number of strong evidence sources across cohorts.

**e,** Fraction of diagnostic genes receiving strong support from the Phenotype, Database, and In-Silico Labs in each cohort.

**f,** Distribution of diagnostic genes by the number of strong evidence sources across cohorts.

**g,** Top-*K* recall curves for RareCollab, AI-MARRVEL, and Exomiser in the UDN, BG, and DDD cohorts.